\documentclass{iaus}
\usepackage{graphicx}

\title[S265~~Turbulent Mixing in Stars] 
{Turbulent Mixing in Stars: Theoretical Hurdles}

\author[W. David Arnett \& Casey Meakin]   
{W. David Arnett
 \and Casey Meakin}

\affiliation{Steward Observatory, University of Arizona, \\ 933 Cherry Avenue,
Tucson, Arizona 85721, USA \\ email: {\tt darnett@as.arizona.edu} 
\\ email: {\tt casey.meakin@gmail.com} }

\pubyear{2009}
\volume{265}  
\pagerange{119--126}
\jname{Chemical Abundances in the Universe: Connecting First Stars to Planets}
\editors{K. Cunha, M. Spite \& B. Barbuy, eds.}
\begin{document}

\maketitle

\begin{abstract}
A program is outlined, and first results described, in which fully three-dimensional, time
dependent simulations of hydrodynamic turbulence are used as a basis for theoretical investigation
of the physics of turbulence.
The inadequacy of the treatment of turbulent convection as a diffusive process is indicated.
A generalization to rotation and magnetohydrodynamics is indicated, as are connections to
simulations of 3D stellar atmospheres.

\end{abstract}

\firstsection 
\section{Introduction}
John von Neumann (\cite[von Neumann 1948]{vonneuman48}) proposed a way to deal with the
intractable problem of hydrodynamic turbulence, by (1) using numerical simulation on computers to
construct turbulent solutions of the hydrodynamic equations, (2) building intuition from study of these
solutions, and (3) constructing analytic theory to describe them. He proposed that iterating this procedure could lead to a practical understanding of turbulent flow. 
The computer power available at that time
was totally inadequate to compute hydrodynamics on sufficiently refined grids to produce turbulent
flow; numerical viscosity restricts the effective Reynolds number. Today, computing power is adequate
for the simulations of truly turbulent, three-dimensional (3D), time dependent, compressible flows,
so we have begun a program based upon von Neuman's proposal.

Turbulent flow in its many guises (e.g., convection, overshooting, shear mixing, semi-convection, etc.) is probably the weakest aspect of our theoretical description of stars (and accretion disks). 
The full problem
that faces us includes rotation, magnetic fields, and multi-fluids (to account for compositional heterogeneity, diffusion, radiative levitation, and nuclear burning). In this paper we describe the
progress made toward von Neumann's goal. We plan to replace the venerable mixing-length
theory (MLT) with a physics-based mathematical theory which can be tested by refined simulations
and terrestrial experiment (e.g., laboratory fluid experiments, meteorological and oceanographic
observations). Particularly relevant are high-energy density plasma (HEDP) experiments, which
now can access regions of temperature and density that overlap stellar conditions up to helium
burning (\cite[Remington et al. 2000, 2006]{}, \cite[Drake 2006]{}), and deal with plasma and magnetic fields, just like star matter, not with an unionized fluid like air or water. 

\section{Inadequacy of the Diffusion Model of Convection}

It is numerically convenient to replace convective mixing in a stellar evolution code by a diffusion algorithm, but this is not physically correct.
The correct equation for the change of composition $Y_i$ is (\cite[Arnett 1996]{}),
\begin{equation}
\partial Y_i + {\bf v \cdot \nabla} Y_i = - {\bf u \cdot \nabla} Y_i + {\cal R}_i,
\end{equation}
where the term on the left-hand side is the Lagrangian time derivative of the composition in
a comoving spherical shell with velocity $\bf v$,
the first term on the right-hand side is the mixing due to rotation and turbulent velocities $\bf u$ across
the Lagrangian shell boundary, and the last term is the composition change due to nuclear
reactions which change species~$i$. Thus, 
\begin{eqnarray}\nonumber
{\cal R}_i =  - Y_i \Lambda_i  + Y_j \Lambda_j \\
 - Y_i Y_k \rho {\cal N}_A \langle \sigma v \rangle + Y_l Y_m \rho {\cal N}_A \langle \sigma v \rangle
 + \cdots
\end{eqnarray}
where the terms on the right-hand side represent all the ways in which species $i$ can be made or destroyed.
The advection operator 
\begin{equation}
{\bf - u \cdot \nabla } Y_i
\end{equation}
 involves a velocity field $u$ which is determined
non-locally and a first order spatial gradient ${\bf \nabla} Y_i$. This is replaced by 
\begin{equation}
 {\partial \over \partial m}[(4 \pi r^2 \rho)^2 D ({\partial Y_i \over \partial m})],
 \end{equation} 
 which has a second order derivative in space and a phenomenological local diffusion coefficient $D$. Except for contrived cases, these are the same
(zero) only in the limit that composition is homogeneous. 
We need a major community effort to base stellar mixing algorithms on physics, comparable to the efforts led by Willy Fowler for nuclear reaction rates, so that {\em both} the advection and reaction terms are reliable.

\section{The Simulation Step}

We have simulated turbulent flow resulting from shell oxygen burning in a presupernova star.
Because of the fast thermalization time (unlike the solar convection zone, for example) we can
simulate the entire convective depth as well as the stable boundaries.
This is a "convection in a box" approach,  implicit large eddy simulation (ILES). 
Using a monotonicty preserving treatment of shocks (like PPM,  see \cite[Boris 2007]{} and
\cite[Woodward 2007]{}) 
insures that the turbulent energy
moves from large scales to small in a way close to that envisaged by \cite[Kolmogorov 1941]{}, 1962.
Because the rate of the cascade of turbulent flow from large scales to small is set by the largest
scales, there is no need to resolve the smallest scales, which are far below our grid resolution.
This would not be the case if the nuclear burning time were shorter than the turnover time, instead
of a thousand times longer.  A detonation or deflagration, in which the turnover time is much longer
than the reaction time, is  a more difficult problem.

The aspect ratio is chosen to be large enough so that it has little effect on the simulation.
The initial state is mapped from a 1D model with sufficient care so that there is very little transient "jitter".  The convection develops from numerical roundoff noise or from low amplitude seed noise.
A quasi-steady state, in an average sense, develops in one turnover time, so that memory of
initial errors is quickly lost.

The simulations show that this oxygen shell burning is unstable to nuclearly-energized pulsations
(primarily radial), which couple to the turbulent convective flow. The convective kinetic energy
shows a series of pulses with an amplitude change of order of a factor of two. 
These disappear if the burning is
artificially turned off. For more detail, see \cite[Meakin \& Arnett (2007b)]{}. 
\cite[Meakin \& Arnett (2006)]{} find that 2D simulations which include multiple burning shells show
interactions between the shells; 3D simulations of multiple shells are planned.
Neither have the pulsations, nor the interaction of burning shells, been included in
any 1D progenitor models to date. 

Another novel feature found by \cite[Meakin \& Arnett (2007b)]{} is entrainment at convection boundaries. The physics of the process is interesting; it involves the erosion of a stably-stratified
layer by a turbulent velocity field, mediated by nonlinear g-mode waves 
(\cite[Meakin \& Arnett 2007a]{}).

While these simulations do not contain an entire star, and thus limit the accuracy of the description
of low-order modes, whole star simulations are developing enough resolution to exhibit turbulent
flows (\cite[Brun 2009]{}). Since we find that even modest resolution will give reliable average 
quantities (see below), we expect the "simulation step" to be soon generalized and 
extended to include rotation and magnetic fields.

\section{The Analysis Step}

The pressure, density and velocity were subjected to a Reynolds decomposition, in which
average properties and fluctuating properties are separated. For example, for pressure,
$P = P_0 + P'$,  so that averages give $\langle P' \rangle = 0 $ and 
$ \langle P \rangle  = P_0$. Note that in general $ \langle (P')^2\rangle \neq 0.$  We use
two levels of averaging: one over solid angle (the extent of our grid in $\theta$ and $\phi$),
and one over time (two turnover times). The resulting averaged properties have a robust behavior
that was insensitive to grid size, aspect ratio, and limits to the size of the averaging dimension
(provided it was large enough; two turnover times and 60 degrees worked fine).

\cite[Arnett, Meakin, \& Young (2009a)]{} find that the velocity scale is well estimated by equating the
increase in kinetic energy due to buoyant acceleration to the decrease due to turbulent damping
in the Kolmogorov cascade.
This implies that it will be possible to make quantitatively correct estimates of wave generation
and entrainment at convective boundaries. \cite[Arnett, Meakin, \& Young (2009a)]{} have shown
that, in the solar case, the velocity scale is significantly larger than estimated by MLT 
(a factor of $\sim 2$),
but agrees with both 3D atmospheres (\cite[Asplund 2005]{}, \cite[Nordlund \& Stein 2000]{} and \cite[Stein \& Nordlund 1998]{}) and empirical solar surface  models (\cite[Fontenla et al. 2006]{}).
The flow becomes more asymmetric as the depth of a convection zone increases (i.e., the upflows are broader and slower), so that there is
a non-zero flux of turbulent kinetic energy, and for deep convection zones ($\geq 1 H_P$, where
$H_P$ is a pressure scale height), the turbulent energy flux is significant relative to enthalpy flux
($ \sim  0.1$) and oppositely directed.

\section{Future Prospects}

\subsection{Comparison to Stars}
These insights are being implemented into an algorithm for stellar evolution\footnote{ John Lattanzio
has dubbed this the "321D" algorithm.}. The idea is to use fully 3D phenomena, found in simulations and captured by analytic theory, by projecting them onto a 1D geometry, as used in stellar models.
Unlike MLT, the algorithm is nonlocal and time-dependent, not static. It should be applicable to deep,
nearly adiabatic convection without modification. Because it have some time dependence it should
be useful for models of pulsating stars. It will replace "overshooting" and "semi-convection" because
it uses the bulk Richardson criterion for the extent of convection (\cite[Meakin \& Arnett (2007b)]{}).
Because the turnover flow in the convection zone is averaged over, this algorithm is not limited
by the corresponding Courant condition, and is appropriate for stellar evolution over long time scales.
We emphasis that failure is possible now that free parameters are being eliminated, so that
inadequacies of the theory will be evident.

\subsection{3D Hydrodynamic Atmospheres}
The "321D" approach merges naturally with work on 3D atmospheres (\cite[Arnett, Meakin, \& Young 2009]{} and above) and work on accretion disks (\cite[Balbus \& Hawley 1998]{}, \cite[Balbus 2009]{},
\cite[Blackman 2010]{}, and \cite[Blackman \& Pessah 2010]{}). 
These approaches all use mean field equations, starting from the same general equations of
mass, momentum and energy conservation for fluids, and use averaging to derive general
properties. Because of this, physical processes are not introduced in patchwork fashion, but
a logical necessities of the conservation laws. Insights into MHD in disks can spark insight into
angular momentum transport in stars, and insights into stellar turbulence should do the same for
accretion disk theory. As Bohdan Paczynski was fond of saying, "accretion disks are just flat stars."

\begin{figure}[b]
\begin{center}
 \includegraphics[width=3.4in,angle=270]{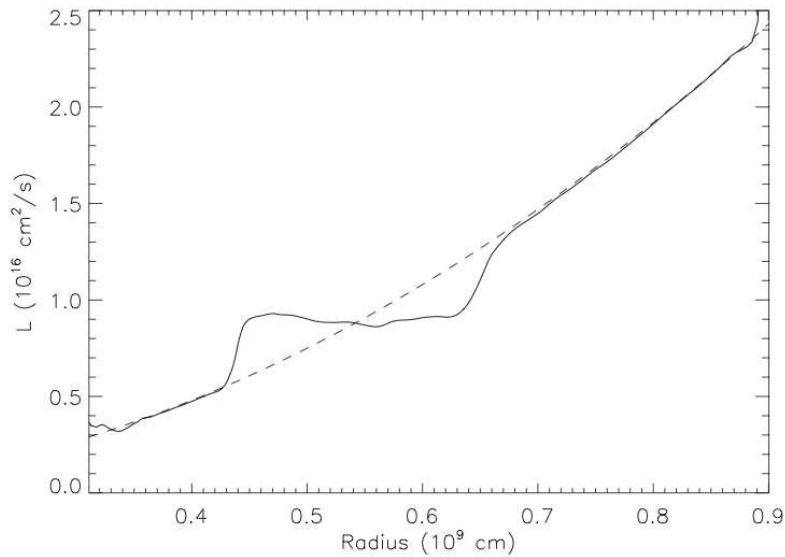} 
 \caption{Specific angular momentum versus radius. The convection zone readjusts to
 constant specific angular momentum, not rigid body rotation (\cite[Meakin \& Arnett 2010]{}).}
   \label{fig1}
\end{center}
\end{figure}

\subsection{Rotation and Magnetic Fields}
Perhaps the greatest challenge for stellar evolution is the treatment of angular momentum
transport. The rigid rotation of the Sun's radiative core, and the differential rotation of the convective
envelope, inferred from helioseismology, seem to have been a surprise. If we wish to understand
GRB's and hypernovae, most workers seem to assume that a key role is played by rotation in
the gravitational collapse and explosion (an idea dating back to Fred Hoyle, at least). We expect
to have little success if we extrapolate from the sun, using algorithms that give the wrong
qualitative behavior. The von Neumann proposal, generalized to include rotation and magnetic
fields, offers hope.

Figure~1 shows the results of a first step toward understanding that problem. Our convection in
a box simulation is continued, but with the box being rotated around the polar axis. The initial
rotation is rigid body, so that the specific angular momentum is quadratic in the radius.
After a few turnover times, the results in Figure~1 is obtained, in which the specific angular
momentum tends toward a constant in the convection zone, while remaining rigid body outside.
Further, magnetic instabilities (MRI, etc.) seem to cause radiative regions to tend toward rigid body
rotation, even if they initially have some other rotation law. A perusal of the literature suggests
that in stellar evolution, the opposite is often assumed. 

\section{Conclusion}
The von Neumann proposal of using computation and theory together seems to work well for
stellar turbulence, and promises to be of value for the more complex problem which includes
rotation and magnetic fields. Perhaps the best aspect of this approach is that it certainly will
make new predictions of phenomena which hitherto have been essentially
in the realm of observation only.

\end{document}